\documentclass[preprint2]{aastex}
\usepackage{graphicx}
\usepackage{epsfig}     
\usepackage{amsmath}
\usepackage{txfonts}
\usepackage{natbib}
\usepackage{multirow}
\bibpunct{(}{)}{;}{a}{}{,}
\usepackage{url}
\newcommand{\changes}{}
\newcommand{\changesa}{}

\slugcomment{}

\shorttitle{Infrared Dark Clouds}
\shortauthors{Marshall et al.}

\begin{document}
  \title{Distribution and characteristics of Infrared Dark Clouds\\using genetic forward modelling}


  \author{D.J. Marshall and G. Joncas} 
  \affil{D\'epartement de physique, de g\'enie physique et d'optique et Centre de recherche en astrophysique du Qu\'ebec, Universit\'e Laval, Qu\'ebec, QC, G1V 0A6, Canada  }
  \and \author{A.P. Jones}
  \affil{Institut d'Astrophysique Spatiale, b\^atiment 121, Universit\'e Paris-XI, Orsay, 91405, France}


\begin{abstract}
  Infrared Dark Clouds (IRDCs) are dark clouds seen in silhouette in mid-infrared surveys. They are thought to be the birthplace of massive stars, yet 
remarkably little information exists on the properties of the population as a whole {\changes (e.g. mass spectrum, spatial distribution)}.
Genetic forward modelling is used along with the Two Micron All Sky Survey and the Besan\c{c}on Galactic model to deduce the three dimensional 
distribution of interstellar extinction towards previously identified IRDC candidates. 
This derived dust distribution can then be used to determine the distance and mass of 
IRDCs, independently of kinematic models of the Milky Way.
 Along a line of sight that crosses an IRDC, the extinction is seen to rise sharply at the distance of the cloud. 
Assuming a dust to gas ratio, the total mass of the cloud can be estimated.

The method has been successfully applied to 1259 IRDCs, including over 1000 for which no distance or mass estimate currently exists.
The IRDCs are seen to lie preferentially along the spiral arms and in the molecular 
ring of the Milky Way, reinforcing the idea that they are the birthplace of massive stars.
Also, their mass spectrum is seen to follow a power law with an index of $-1.75\pm0.06$, 
steeper than giant molecular clouds in the inner Galaxy, but comparable to clumps in GMCs. 
This slope suggests that the IRDCs detected using the present method are not gravitationally bound, 
but are rather the result of density fluctuations induced by turbulence.

\end{abstract}



\section{Introduction}

The first reported detection of Infrared Dark Clouds (IRDCs) was by \cite{Perault1996}, who noted regions 
in mid-infrared (MIR) ISOCAM observations that were optically thick at 15 $\mu$m. \cite{Egan1998} noted that 
many of these dark features do not emit significantly in the far-infrared, implying that they are cold ($<20$ K) and dense ($n>10^5$ cm$^{-3}$) 
molecular clouds.

There has been much interest recently in IRDCs, as they are thought to be the initial phase of massive star formation. 
Indeed, active star formation has been detected in a number of dense IRDCs \citep{Ormel2005, Rathborne2005, Pillai2006, Chambers2009}. 
In order to obtain more information on the clouds themselves as a population, \cite{Simon2006} analysed MIR images from 
the Midcourse Space Experiment \citep[MSX,][]{Price1995}. They searched for contiguous regions of high decremental contrast with respect 
to the MIR background and they identified over 10000 IRDC candidates in the first and fourth Galactic quadrants. In a subsequent study, 
\cite{Simon2006a} used $^{13}$CO observations to determine kinematic distances to and masses of over 300 of the IRDC candidates. 
\cite{Rathborne2006} used the kinematic distances determined by \cite{Simon2006a} along with dust continuum observations and determined 
the masses of nearly 40 IRDCs.
More recently, \cite{Jackson2008} used CS observations to determine the kinematic distances to over 300 4th quadrant IRDCs.
{\changes The typical size of the IRDCs found by the above authors is 5 pc, their average  mass is $\sim 3\times10^{3} M_{\odot}$, 
and their distances 
indicate that they lie preferentially in the molecular ring of the Galaxy\citep{Simon2006a, Jackson2008}. The mass spectra for the IRDCs is found to follow a power law, 
with a spectral index steeper than that found for Giant Molecular Clouds in the inner Galaxy\citep{Simon2006a}. The cores within the IRDCs, however, are steeper 
still and are very close to the Salpeter IMF\citep{Rathborne2006}.}

The area of the Galaxy where 
the detection of IRDCs is easiest, namely inside the solar circle due to the high MIR background, 
is also the region where distance determinations are the most problematic.
Tackling the problem from a variety of angles is therefore very important in order to avoid introducing bias.
In order to extend and complement the latest findings on the Galactic distribution of IRDCs, 
a new tridimensional extinction mapping method is used to locate their positions and line of sight extinctions.
Stars observed in the near infrared are compared to a Galactic Population Synthesis model \citep{Robin2003}, and the extinction 
required to redden the modelled stars to match the observed stars is found via a genetic algorithm \citep[GA,][]{Charbonneau1995}.
Near infrared data are well suited to study these IRDC candidates as $K_s$ band extinction is only a factor of two higher than 
the extinction at 8 ${\mu}$m \citep{Indebetouw2005}.

In section \ref{sec:obs} we describe the observations and model used to calculate the extinction. 
In section  \ref{sec:method} we describe how we determine the line of sight extinction to IRDC candidates, and how this information is 
converted to a distance and mass estimate for the cloud. We describe the results in section \ref{sec:results} and discuss the implications 
in section \ref{sec:discussion}. We conclude in section  \ref{sec:conclusion}.

\section{Model and observations}
\label{sec:obs}
\subsection{Two Micron All Sky Survey}

The Two Micron All Sky Survey  \citep[2MASS,][]{Skrutskie2006} is a ground based survey which
uniformly scanned the entire sky in three near-infrared bands ($J$, $H$ \& $K_{s}$). 
Amongst its final products is the point source catalogue (PSC) which includes 
point sources brighter than about 1 mJy in each band, 
with signal-to-noise ratio (SNR) greater than 10, using a pixel size of 2.0\arcsec. 
{\changes It is complete down to a limiting magnitude of 15.7, 15.1 and 14.3 in the 
$J$, $H$ and $K_s$ bands, respectively, and in the absence of confusion.}

\subsection{The Galactic model}
The stellar population synthesis model of the Galaxy constructed in Besan\c{c}on \citep{Robin2003},
hereafter called the Galactic model,
  is able to simulate
the stellar content of the Galaxy by modelling four distinct stellar populations:
the thin disc, the thick disc, the outer bulge and the spheroid. It also takes
into account the 
dark halo and a diffuse component of the interstellar medium. It can be used to generate 
stellar catalogues for any given direction, and 
returns information on each star such as
magnitude, colour, and distance as well as kinematics and other stellar parameters.

The approach of the Galactic model is semi-empirical as it is based on 
theoretical grounds (for example stellar evolution, galactic evolution and galactic dynamics) and is constrained
 by empirical observations (the local luminosity function for example). 
The Galactic potential is calculated in order
to self-consistently constrain the disc scale height. In addition, the model {\changes simulates photometric errors}
and {\changes includes} Poisson noise to make it ideal for direct comparison with observations.

The Galactic model has been used for various studies, such as identification of Galactic structures 
through stellar overdensities \citep{Picaud2003,Bellazzini2004,Momany2004}, Galactic parameter estimation 
from fitting model to observations \citep{Picaud2004, Reyle2009} as well as the
 quantification of the uncertainty in extinction estimation \citep{Froebrich2006} to name but a few. 
Many more examples are given in \cite{Robin2003}.

\subsection{Comparing model to observations}

The Galactic model has been developed 
to return results in the near-infrared and visible filters, and is a powerful tool to extract the extinction 
information embedded in the 2MASS observations \citep{Marshall2006}.

However, the PSC can be compared quantitatively with the Galactic model simulations only where the former
is shown to be complete. As such, we must compute the faint completeness limit field by field. 
{\changes We do so by creating magnitude histograms in each band and locating the faintest magnitude where the 
 source counts are still increasing linearly.} We use 
a bright cutoff of 10 in all three bands, as recommended in the online explanatory supplement\footnote{\url{http://www.ipac.caltech.edu/2mass/releases/allsky/doc/sec6_5a1.html}}.
In order to closely model the 2MASS stars, we cut the modelled stars at the 2MASS completeness limits and 
simulate the photometric errors via an exponential function which is able to 
reproduce typical CCD photometric errors very well.

\section{Method}
\label{sec:method}
The method used to determine the distance and extinction to the IRDCs is based on  a modified version of the three dimensional extinction method 
first described in \cite{Marshall2006}. The base of the extinction method remains a comparison of the stellar colours 
of the Galactic model and the 2MASS dataset, {\changes as extinction gives rise to a colour excess. For example, for the $J$ and $K_s$ bands the colour excess is:}
\begin{equation}
  E(J-K_s) = (J-K_s) - (J-K_s)_0
\end{equation} 
{\changes where the $0$ subscript denotes intrinsic colour. This equation can be rewritten as}:
\begin{equation}
  E(J-K_s) = (J-J_0) - (K_s - {K_s}_0) = A_J - A_{K_s}
\end{equation} 
{\changes as the extinction in a given band is just the difference between observed and unredenned apparent magnitude}.
Putting these relations together, we can write:
\begin{equation}
\label{eqn:ext}
  A_{Ks} = R_{JK} \, E(J-K_s)
\end{equation} 
{\changes where $R_{JK} = (1/(A_J/A_{Ks} -1))$ is the ratio of absolute to relative extinction.}
 The modifications to the basic method are described below.

\subsection{Genetic algorithm approach}

In \cite{Marshall2006} the authors calculate the extinction via the colour difference between observed reddened stars and 
simulated unreddened ones {\changes(i.e. to which no extinction correction has been applied)}. 
The distance information is then obtained by assuming that more distant stars are redder, as 
interstellar extinction in a monotonically growing function with distance. This is true for the giant star population only, 
so dwarf stars are excluded from the calculation.

An alternative is to turn the problem on its head. Instead of calculating an extinction based on colour differences between 
observed and modelled stars, one can {\changes generate} a large number of extinction distributions and keep the one that results in a good 
fit between observed and modelled stars. An immediate advantage of this method is that the information from all of the stars 
is used, not just the giants.
However the sheer number of possible solutions prevents us from applying a ``brute force'' 
approach. We have instead applied the genetic algorithm (GA) approach described in \cite{Charbonneau1995}.
\begin{figure*}[t]
  \begin{center}
    \includegraphics[width=8cm]{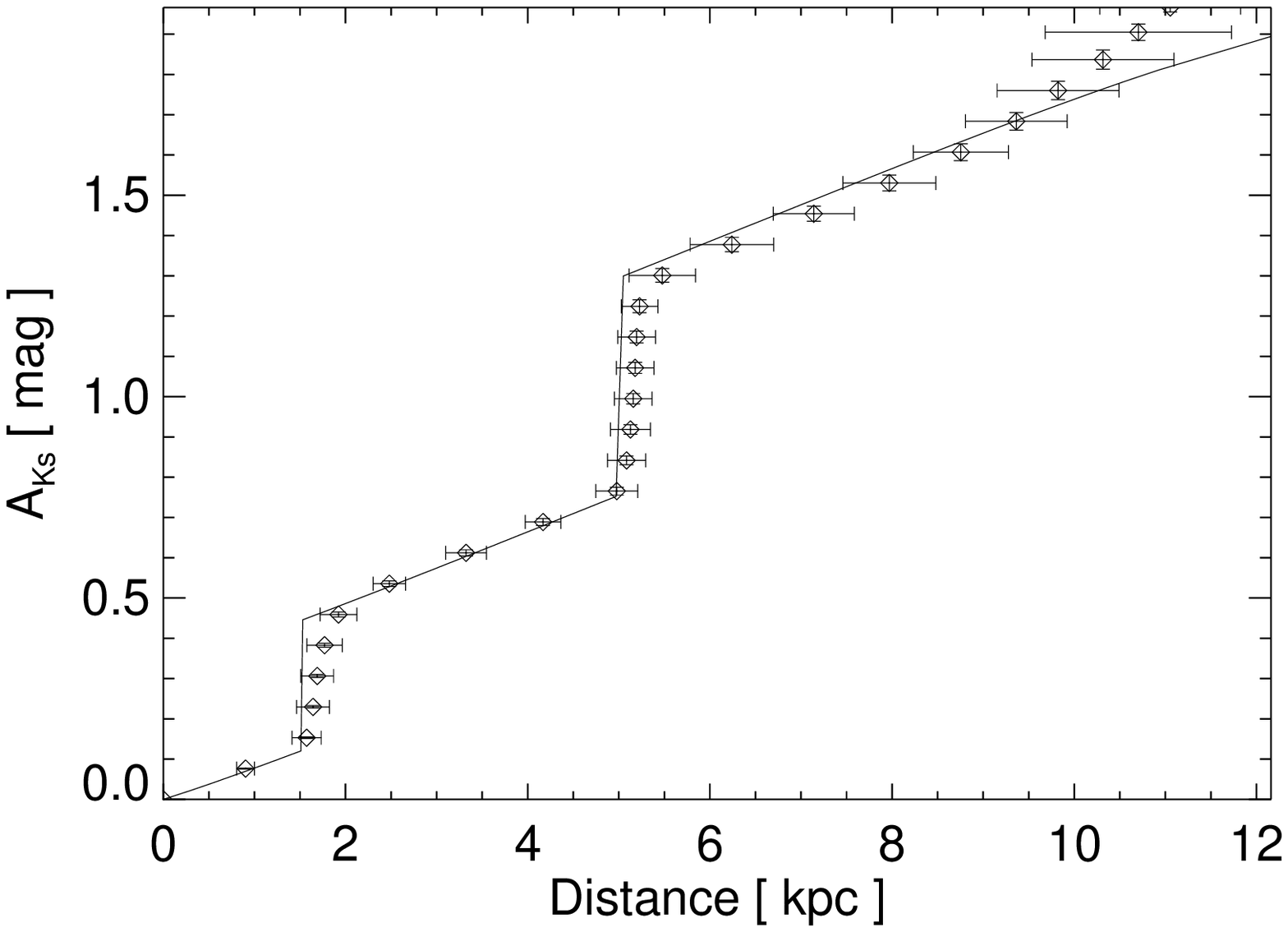}
    \includegraphics[width=8cm]{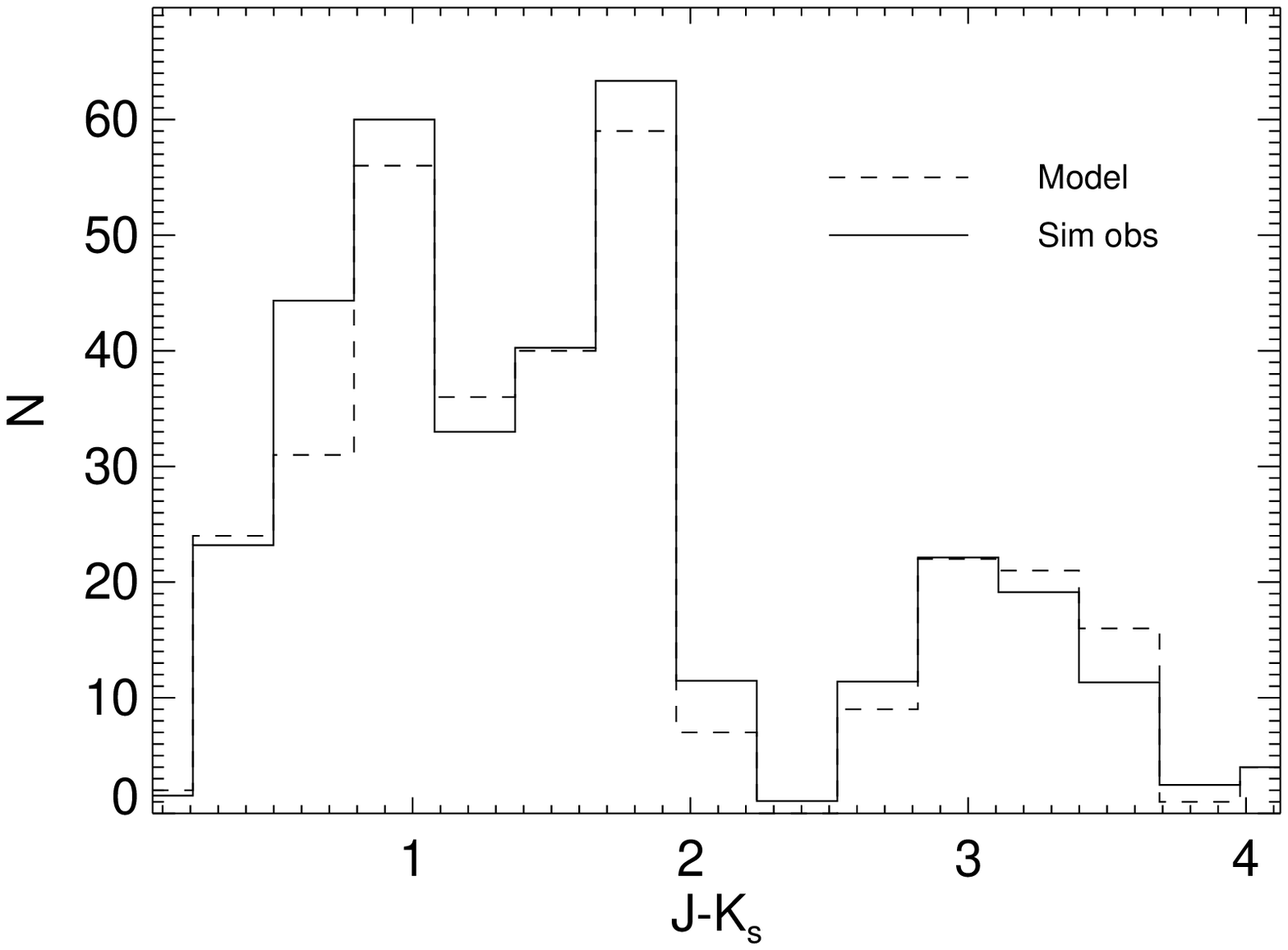}
    \caption{{\changes {\bf Left:} Extinction distribution for an example line of sight. The observations have been simulated 
      using the Galaxy model with known extinction distribution (solid line).  Note that the distance error bars 
      generally increase with distance. {\bf Right:} Colour histogram for the $J-K_s$ colour index for the simulated observations (solid line) 
      and the model adjusted with the extinction points of figure on the left.}}
  \label{fig:example}
  \end{center}
\end{figure*}

In essence the GA {\changes generates and} tests different extinction distributions {\changes towards a candidate IRDC},  defined as a number of points each with 
a distance and a corresponding extinction. {\changes The extinction is thus no longer calculated, but is an output from the GA.}
{\changes For each generated extinction distribution, the modelled stars are reddened using Eq.\ref{eqn:ext},}
 and the goodness of fit between the observed and simulated stellar colour distributions {\changes is calculated}
via a {\changes merit} function, described below. Initially, a ``population'' of random solutions is generated.
Solutions in this group with the highest values {\changes of the merit function} are combined to produce 
``child'' solutions in a new ``generation'' of solutions. This evolution is allowed to continue for at most 500 generations 
after which the fittest solution found is retained.
If the solution converges before then, defined as 25 generations with no improvement in the fittest individual, 
the iteration is stopped and the fittest solution {\changes found up to that point} is adopted.

The use of Galactic modelling and GAs to solve for three dimensional extinction is an interesting alternative 
to {\changes other} techniques, as no single stellar population need be isolated, and there is no such thing as foreground 
contamination - all stars are used in the extinction calculation.
However these advantage comes at a price - high CPU use resulting in a long execution time. 
Each point along a line of sight {\changes towards the IRDC, and contained within its defining ellipse,} 
corresponds to two parameters (distance and extinction) and so 
by reducing the number of parameters to fit, a substantial gain in time is possible. This is done by 
fixing the extinction values on a regular grid along each line of sight
 and searching only for the corresponding distance to each extinction value.
This approach is superior to fixing the distance to a regular grid, as sharp rises in extinction 
are easier to identify.

{\changes The extinction values along the line of sight are thus equally spaced:}
\begin{equation}
  A_{Ks}(i) = i \times \delta A_{Ks} 
\end{equation}
for each line of sight point $i$ and where $\delta A_{Ks}$ is taken to be twice the minimum extinction detectable. 
Assuming that the minimum extinction detectable 
arises from the uncertainty in the 2MASS $J-K_s$ colour, this value is then{\changes, using Eq.\ref{eqn:ext}}:
\begin{equation}
  \delta A_{Ks} = 2 * \frac{\sqrt{(\Delta J^2 +\Delta K_s^2)}}{(A_J/A_{Ks})-1}
\end{equation}
where $\Delta J$ and $\Delta K_s$ are the average uncertainty in the magnitude of the $J$ and $K_s$ bands, respectively.
{\changes Typically, this corresponds to $0.06 \la \delta A_{Ks} \la 0.1$.}
The distribution of photometric errors is Poissonian, requiring the use of the mean in the above equation so as 
to not allow the extinction calculation to work purely on noise. Also, 
the use of the $J-K_s$ colour index is justifiable as it is the most sensitive of the three to interstellar extinction.

The distance values along the line of sight are:
\begin{equation}
  \label{eqn:deltar}
  r_i = r_{i-1} + \delta r_i \hspace{5mm} i>1
\end{equation}
where $r_1 = 0$ and the $\delta r_i$ are the variables that are searched for using the GA.

We have added two modifications to this schema. First of all, a crude estimate to the solution is estimated by finding a smooth extinction solution.
The Galactic dust distribution in the model is approximated  using a double
exponential disc with an {\it ad hoc} local normalisation. The best local normalisation is found in order to minimise the $\chi^2$ difference 
between observed and modelled stellar colour distributions for the particular line of sight. 
This rough solution is inserted into the initial population as an ``alpha male'', intended 
to get the GA looking in the right direction. 
This ``alpha male'' is generally replaced by a fitter solution within the first couple of generations and does not influence the final solution, but greatly reduces the convergence time. If the initial random generation of solutions contains a fitter individual than the alpha male is not used.

Secondly we have modified one of the reproduction functions used to create child solutions, namely the crossover operator. 
GAs encode the solutions into individual strings called genotypes. 
In our case the solution is a series of $\delta r_i$ values (Eq.\ref{eqn:deltar}) which the GA normalises. 
A simple example consisting of two points which we describe to three decimal digits would then be a six digit number. The crossover operator 
would take two parent genotypes, randomly select a digit to perform the cut, and then simply swap the remaining digits between the two parents to create 
two new children. This singular cut results in a slower mixing of the genotypes so we have changed it to N-point crossover. In this scheme, the number 
of cuts is a random variable and so the children contain more diversity than in the simple crossover operator. This approach emphasises exploration of the 
parameter space, with less exploitation of the fittest individual. We are not looking for a  high precision solution as the colour distribution 
of the observations can be reproduced with a spread of extinction solutions (see Sect.\ref{sec:errors}). Rather we want to avoid any secondary, local minima in the $\chi^2$ minimisation. 


\begin{figure*}[t]
  \begin{center}
    \includegraphics[width=0.495\linewidth]{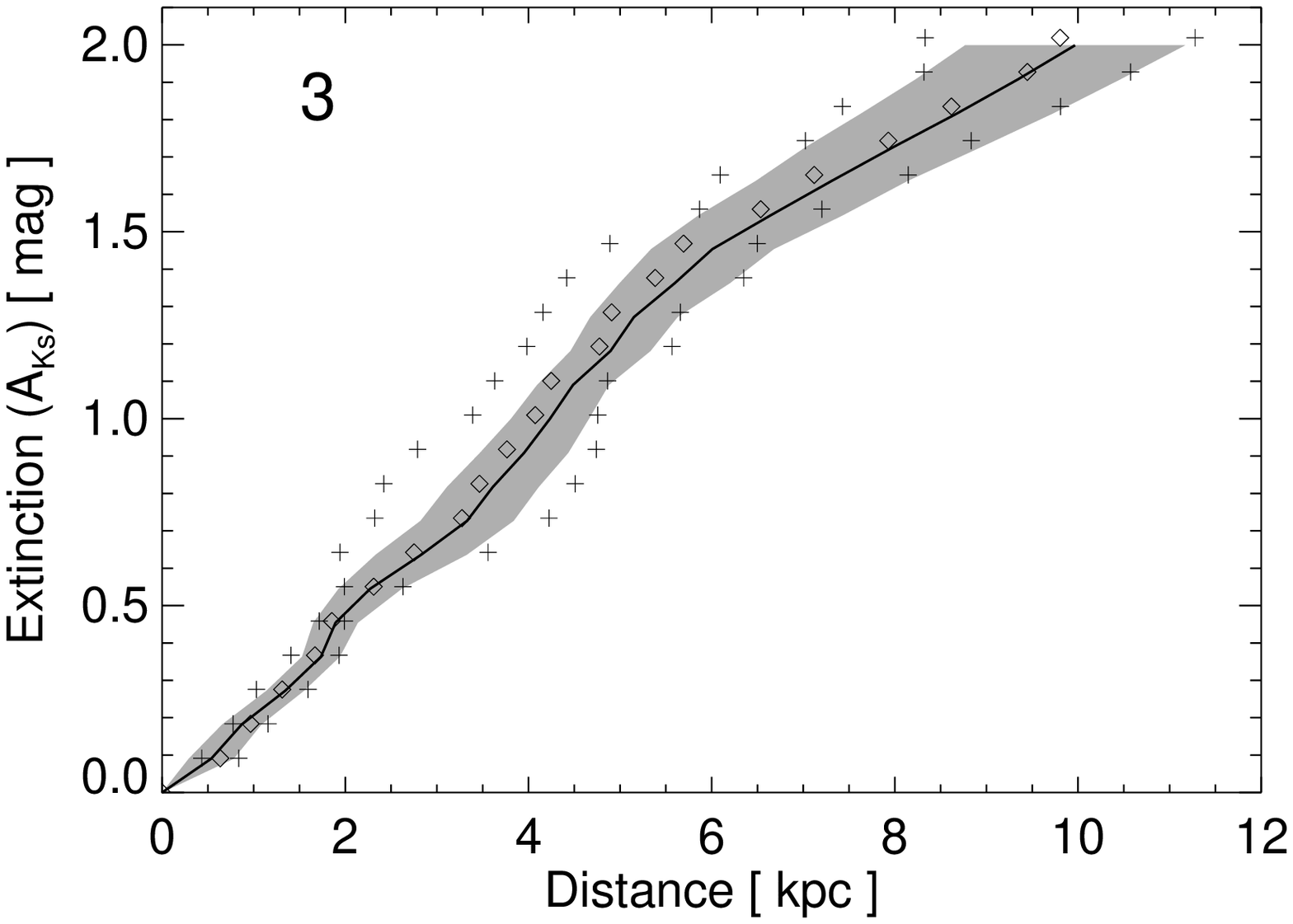}
    \includegraphics[width=0.495\linewidth]{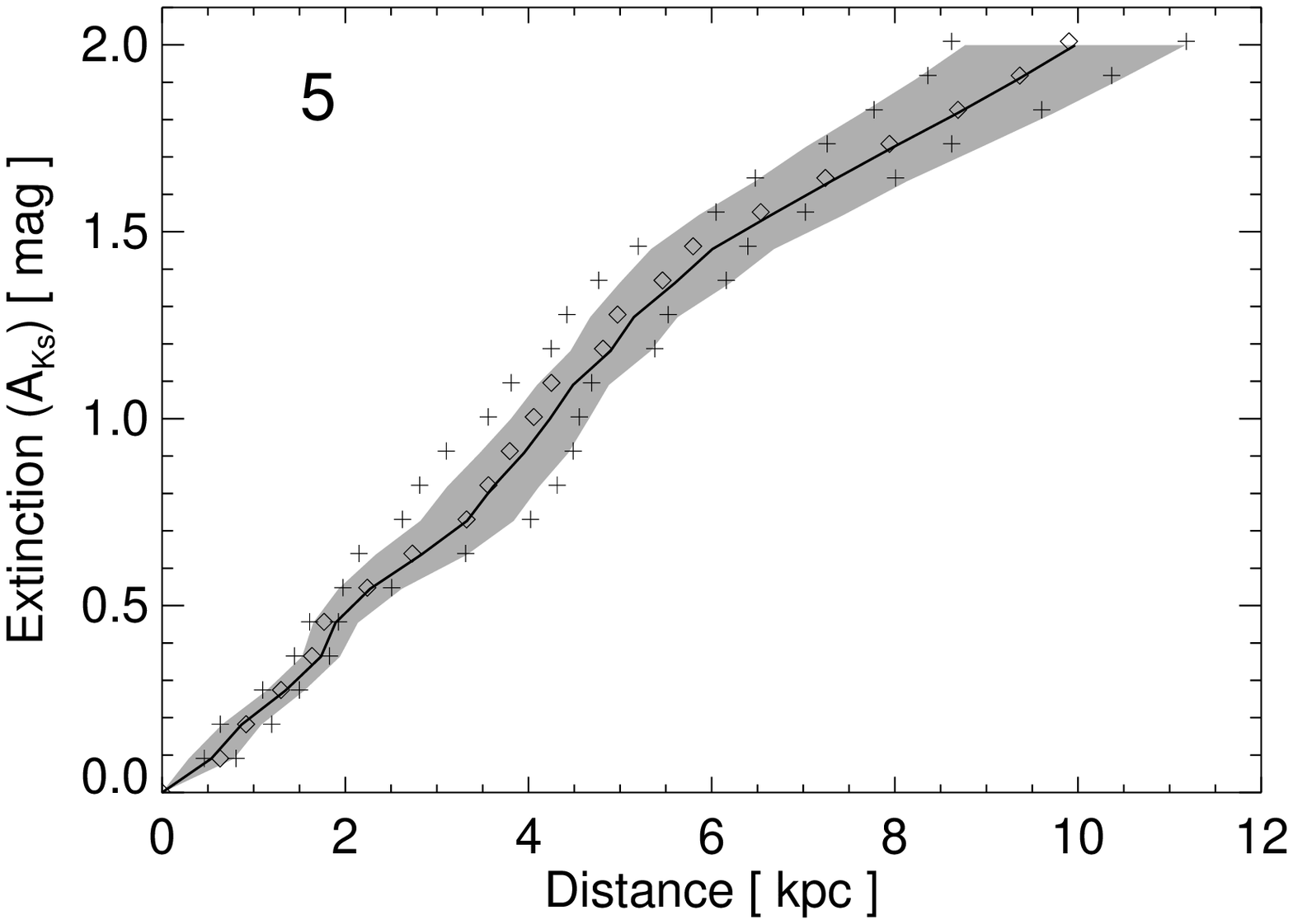}
    \includegraphics[width=0.495\linewidth]{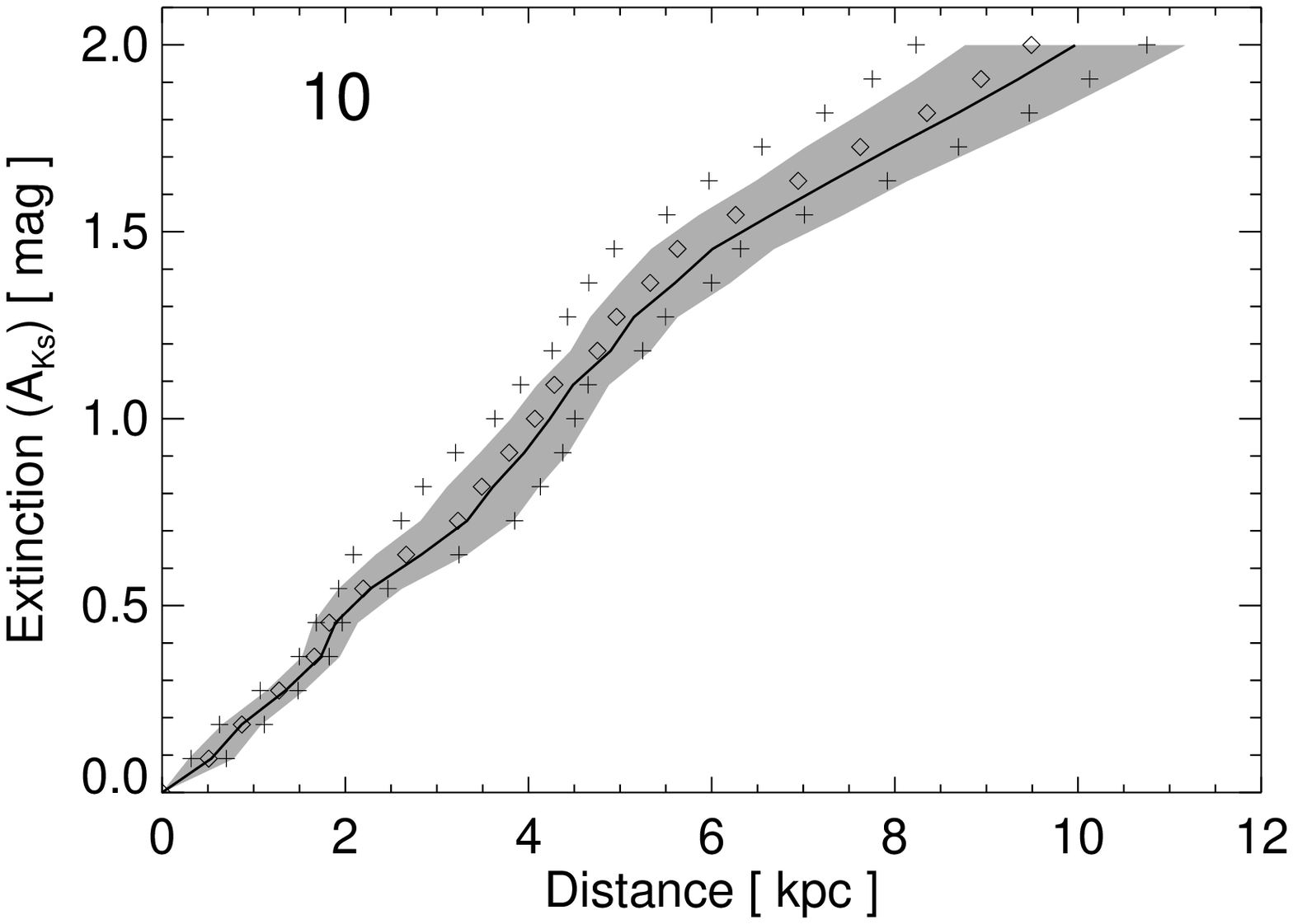}
    \includegraphics[width=0.495\linewidth]{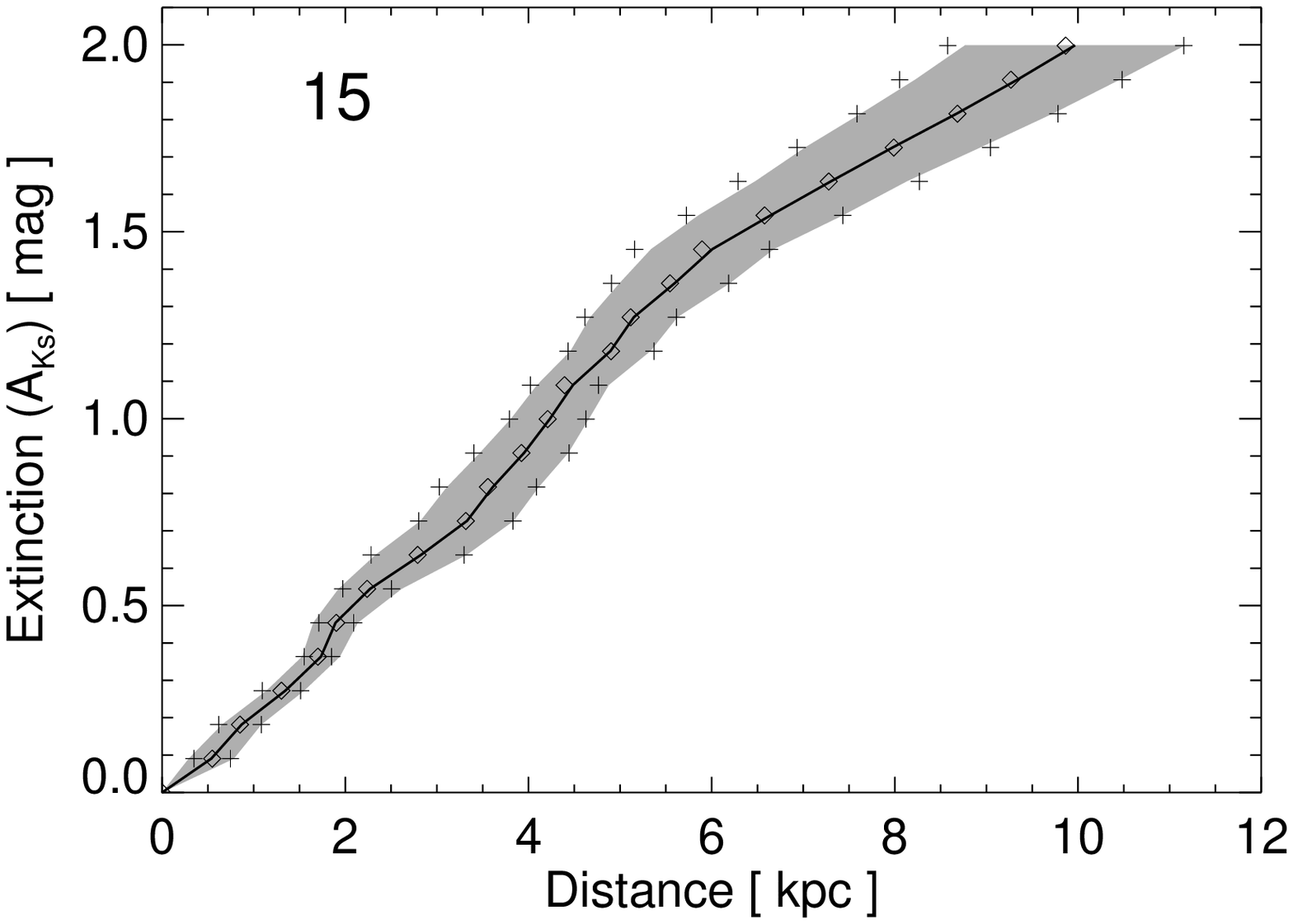}
    \caption{Extinction along a test line of sight at $l=50\degr$ $b=0\degr$. Each figure shows the resulting 
    extinction after averaging a different number of calculations. The solid line and grey shaded area show 
    the average and standard deviation after 100 runs. The diamonds show the mean and the crosses show the 
    standard deviation after the number of runs showed in the top left corner.}
  \label{fig:test_plot}
  \end{center}
\end{figure*}

\subsection{Merit function}

For each line of sight we are comparing observed stars with simulated stars from the Galactic model via the {\changes histograms of their 
respective colour indices ($J-K_s$ and $H-K_s$)}.
We calculate their difference using a $\chi^2$ statistic, {\changes where higher merit is associated with a lower value of the $\chi^2$ statistic}. 
However, if the Galactic model overestimates 
or underestimates the number of stars along a particular line of sight we do not want this to influence the extinction we measure. Consider 
a hypothetical case where the model predicts a factor of two too many stars. If we were to use a standard $\chi^2$ test between modelled 
and observed colour distribution, then the extinction we find could be artificially boosted in order to remove a number of simulated 
stars by pushing them beyond the completeness limit for the line of sight. This is obviously an unacceptable situation. 

To avoid this ``artificial extinction'' the merit function used is based on a normalised $\chi^2$ test from \cite{Press1992} :
\begin{equation}
\label{eqn:X2}
\chi^2 = \sum_i \frac{(\sqrt{N_{\rm o}/N_{\rm m}} \, n_{{\rm m}_i} - 
\sqrt{N_{\rm m}/N_{\rm o}} \, n_{{\rm o}_i})^2}{n_{{\rm m}_i}+n_{{\rm o}_i}}
\end{equation} 
where $n_{{\rm o}_i}$ ($N_{\rm o}$) and $n_{{\rm m}_i}$ ($N_{\rm m}$) 
are the number of stars in the $i^{\rm{th}}$ bin of the colour histogram 
(total number of stars along the line of sight) of the observations and model, respectively.  
The term involving the total number of stars ensures that 
the $\chi^2$ statistic will be lowest when the shapes of the two 
histograms are the same, regardless of any difference in the total number of stars.
To ensure the best fit, {\changes the $J-K_s$ and $H-K_s$ colour} indices are adjusted simultaneously. If only one is used, the GA will find 
the optimum solution for that colour index, perhaps at the expense of one of the others. 
An example line of sight is presented in {\changes of the left hand side of} Fig.\ref{fig:example} using simulated observations from the Galactic model. 
{\changes The simulated observations contain a diffuse extinction component  (of $0.7\, A_V\, {\rm kpc}^{-1}$) along with two 
clouds at 1.5 and 5 kpc with extinctions of 0.34 and 0.57 magnitudes in the $K_s$ band respectively. The stellar density and extinctions used in the simulation 
is typical of that seen for the IRDCs. The histograms of the $J-K_s$ colour index are shown on the right side of Fig.\ref{fig:example} for the simulated observations 
(solid line) and the Galactic model reddened using the points shown on the left hand side of Fig.\ref{fig:example}.}
 The GA is seen to find the diffuse extinction as well as both clouds.
The detection of the clouds and 
the determination of their characteristics will be discussed in section \ref{sect:findcloud} below.

\subsection{Error estimation}
\label{sec:errors}
The random nature of the GA means that two subsequent calculations
 along the same line of sight will not be identical. Furthermore, the Besan\c{c}on 
 model generates simulated stars using 
density distributions and luminosity functions, meaning that generating a simulated stellar catalogue twice 
for the same line of sight will not result in the same exact solution either. However, 
by rerunning the Galactic model and 
repeating the extinction calculation a number of times and taking the mean, a more robust estimator of the extinction 
distribution can be obtained. 
The variation in the solutions found provide us with a measure of the uncertainty, as they provide a range 
of extinction solutions which are able to fit the model to the observations.

Each iteration to calculate the extinction along the line of sight 
 is a lengthy process and we wish to minimise the number of iterations while still performing enough to obtain a robust mean.
In order to determine the number of solutions which should be generated, we calculated the extinction for a few lines of sight 300 times. 
It was found that there was no difference in the mean after 100 and after 300 iterations, so we assume that the method has converged at 100 
iterations and that 
we have found the best solution and associated uncertainty. 
This number of iterations would require a very lengthy calculation for the thousands of candidate IRDCs so 
we searched for the minimum number of iterations which satisfactorily reproduced the result found after 100 iterations. 
The results of this test for one line of sight is 
presented in Fig.\ref{fig:test_plot}. 
The black solid line with the grey shaded area represent the mean and the standard deviation, respectively, after 100 iterations. 
From left to right, and top to bottom we display the results after 3, 5, 10 and 15 iterations. 
As can be seen, there is far too much variation in the mean for iterations less than 15. After 15 iterations, 
the mean and standard deviations are very close to those obtained after 100 iterations. We thus proceed using only 15 iterations.

\subsection{Cloud identification and properties}
\label{sect:findcloud}

\begin{figure}[ht!]
  \begin{center}
    \includegraphics[width=7cm]{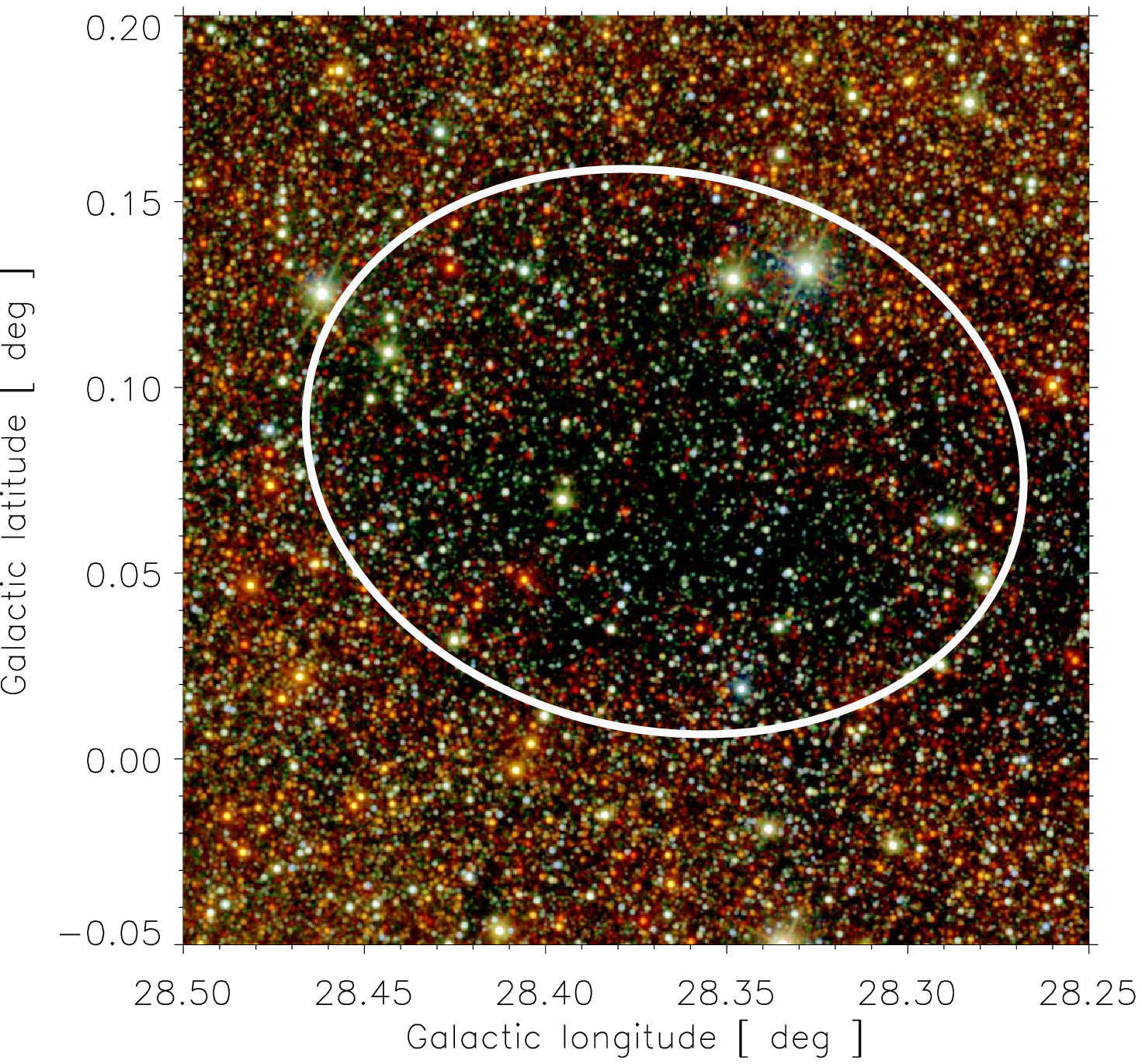}
    \includegraphics[width=7cm]{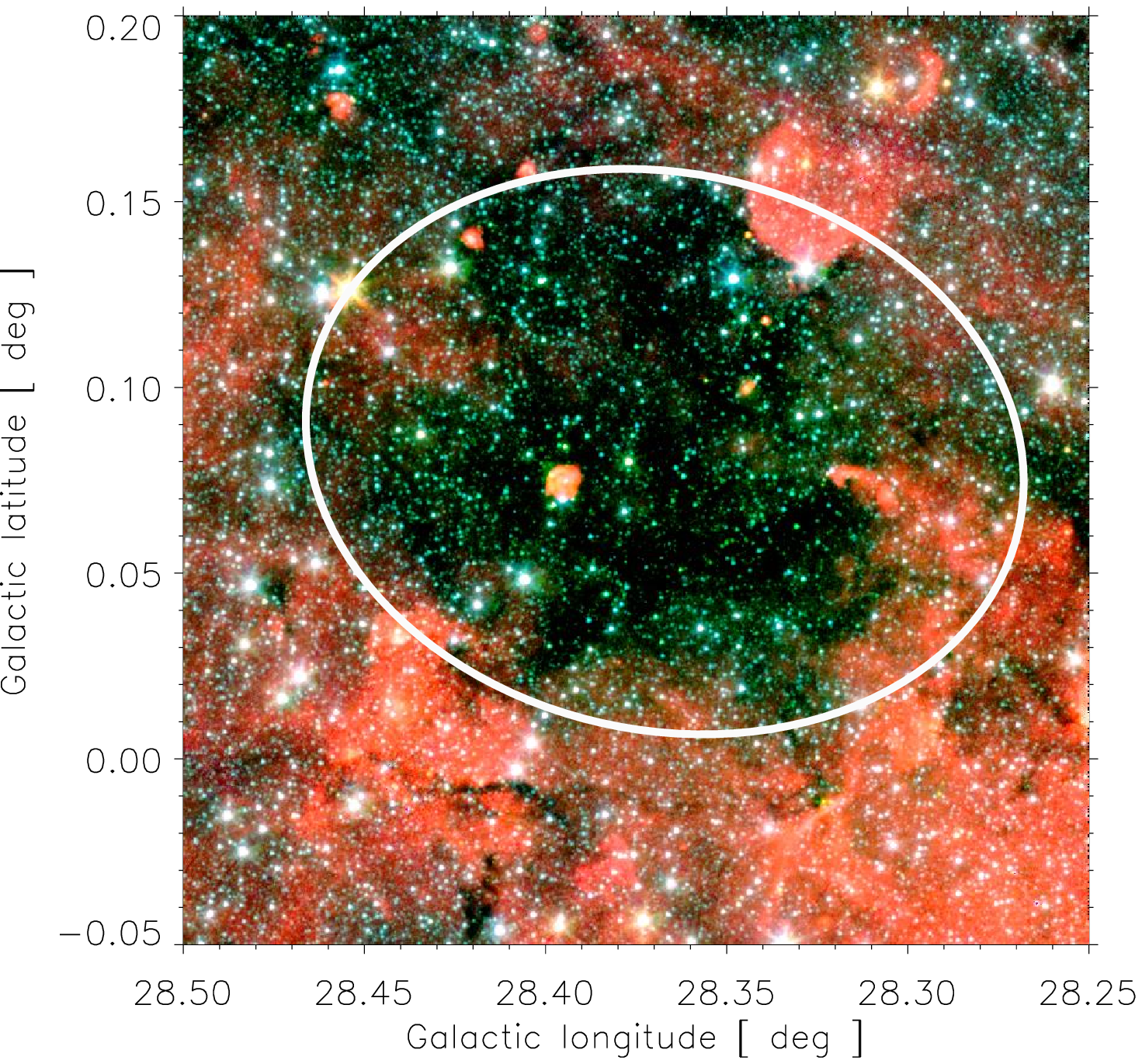}
    \caption{Infrared observations towards MSXDC G028.37+00.07 (ellipse). 
      {\bf Top}: Near infrared observations from 2MASS.
      {\bf Bottom}: Mid infrared observations from Spitzer / GLIMPSE.}
    \label{fig:2mass_glimpse}
\end{center}
\end{figure}
Clouds for which we attempt to measure their extinction come from 
 a catalogue of IRDC candidates from \citep{Simon2006}. In this catalogue the IRDCs are defined as contiguous regions of 
MIR extinction. The catalogue defines ellipses enclosing the clouds and supplies their position, major and minor axes, 
position angle as well as peak contrast and area. For each cloud we fetch the 2MASS observations within the ellipse. All clouds which contain 
over 50 observed stars in $J$, $H$ and $K_s$ are retained. We require all three magnitudes 
 in order to calculate the various  colour indices. 
There is a large spread in stellar colours
towards the dark clouds and we require at least five stars in each colour histogram bin to calculate a meaningful 
$\chi^2$ statistic.
The lower bound of 50 stars is chosen somewhat arbitrarily in order to 
ensure the colour histograms have a sufficient number of stars per bin. 
The line of sight extinction towards this ellipse is then calculated. {\changes In Fig.\ref{fig:2mass_glimpse}, 
an example ellipse from the \cite{Simon2006} catalogue is overplotted on 2MASS data (top) and GLIMPSE mid infrared data (bottom). 
The number of 2MASS stars detected is lower within the ellipse, and the colours of the stars are limited to blue (foreground) and 
heavily reddened (background). The GLIMPSE data shows a substantial decrease of the 8 $\mu$ diffuse emission within the ellipse.
However the number of point source detections, within completeness limits, is nearly three times higher  in the 3.6 ${\mu}$m and 4.5 ${\mu}$m IRAC bands than in the three 2MASS bands. This opens up the possibility of using these longer wavelength observations 
to conduct a similar study in the future.}

\begin{figure}[t]
  \begin{center}
    \includegraphics[width=8cm]{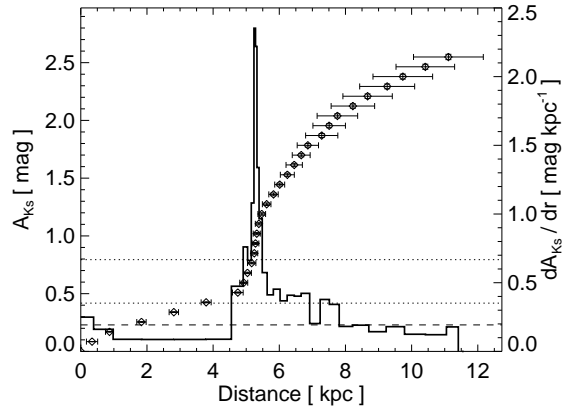}
      \caption{Example of cloud detection for MSXDC G028.37+00.07.
	The diamonds and error bars are as for Fig.\ref{fig:example}. The solid line is the density of absorbing matter 
      ($dA_{Ks}/dr$), the dashed line is the diffuse extinction and the dotted lines show {\changes the diffuse component} $+$ 1 and 3 $\sigma$. 
	A cloud has been detected at $\sim 5.3$ kpc {\bf with an extinction of $\sim 0.9$ mag in the $K_s$ band}.}
    \label{fig:findcloud}
\end{center}
\end{figure}

To determine the extinction of and distance to an individual cloud, we calculate the derivative of the extinction with respect to 
distance along the line of sight (i.e. $dA_{KS} / dr$).
This, in effect, gives us the space density of absorbing matter which in our case is dominated by large dust grains (sizes
typically of the order of 1 $\mu$m). {\changes The value of $dA_{KS} / dr$ can be converted to $n_H$ ({\changesa in total hydrogen, in atomic or molecular form, per cubic centimetre}) by assuming 
a conversion factor between extinction and hydrogen column density. However as this requires us to assume such a value, and as it is just a multiplicative factor, 
we continue to work in units of mag kpc$^{-1}$ in the $K_s$ band. }

{\changes The line of sight extinction will be the sum of a diffuse extinction upon which is superimposed one or more 
clouds. To simplify, we modelize this diffuse component as a constant extinction per unit distance, 
which we choose to be that which minimises the mean absolute deviation (which 
we will denote $\sigma$ in the following) of the line of sight density from the diffuse component. 
As clouds are present along the line of sight, using all the points will of course overestimate the diffuse component. We therefore 
repeat the calculation by omitting all points with a density above $3\sigma$. This is repeated until the fractional change of 
diffuse component and $\sigma$ are less than $2\%$.}

\begin{figure*}[t!]
  \centering
    \includegraphics[width=\linewidth]{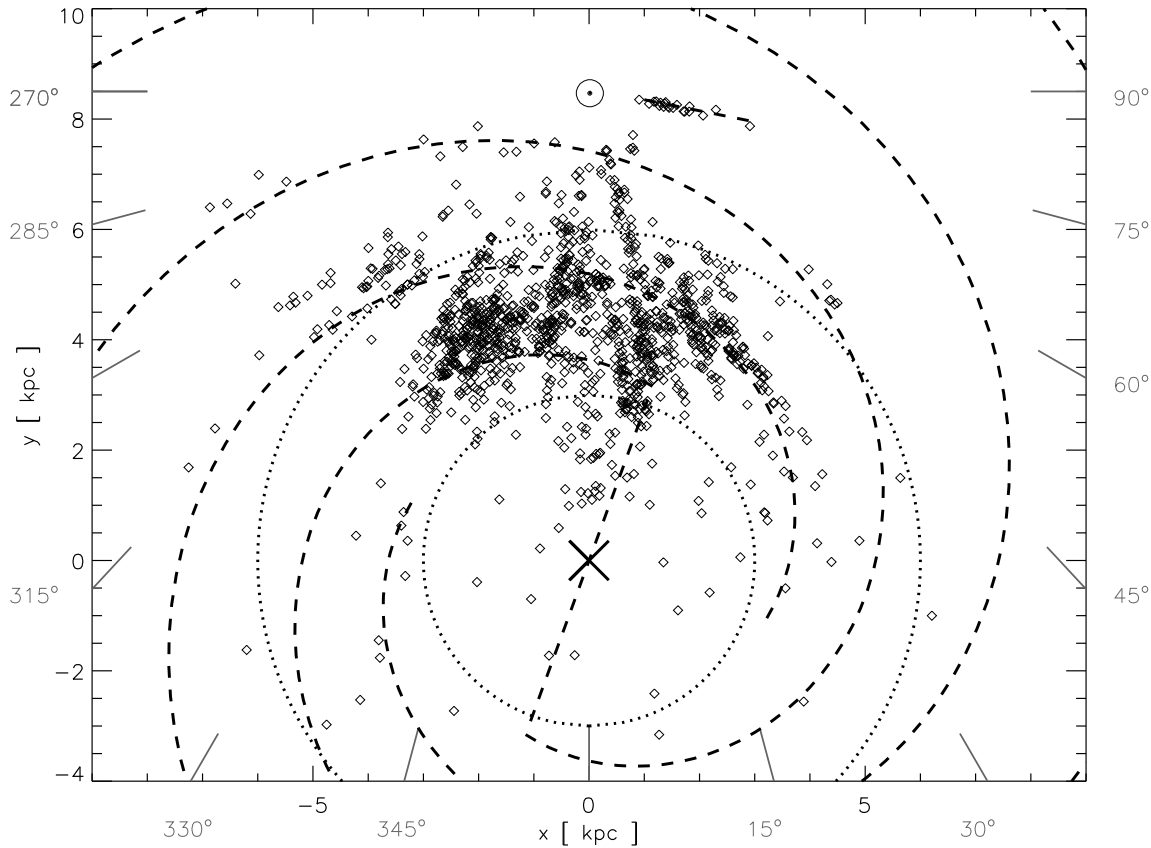}
    \caption{Distribution of the IRDCs in the plane of the Galaxy.  The dashed lines  are the spiral arms from \cite{Vallee2008}, the local Orion spur 
      and the Galactic bar. The  dotted lines show the Galactocentric distances at 3 and 6 kpc.
      The IRDCs are seen to be concentrated in the molecular ring ($3<R<6$ kpc) and along the spiral arms.}
    \label{fig:irdc_pos}
\end{figure*}

All peaks in $dA_{KS} / dr$ over 3$\sigma$ above the {\changes diffuse component} are then considered as a cloud. 
The extent of the cloud is found by finding the adjoining points which remain $1\sigma$ above the {\changes diffuse component.
 The extinction of the cloud is then the increase in the extinction over the cloud minus the diffuse component.}
Only clouds with an extinction over $A_{Ks}\ga 0.114$ are considered, as this threshold 
corresponds to the density necessary for clouds to be dominated by molecular gas \citep[i.e. $A_V\ga 1$][]{Spitzer1973,Binney1998}, 
{\changes if we assume the extinction law of \cite{Cardelli1989}}.
If more than one cloud is found along the line of sight, the one with the largest average 
extinction is retained as the IRDC. An example line of sight is presented in Fig.\ref{fig:findcloud}.

Once identified, the distance to the IRDC is simply the density weighted average over the cloud peak:
\begin{equation}
  r_{\rm IRDC} = \frac{ \sum_{i=i_{\rm min}}^{i_{\rm max}}r_i \times (dA_{Ks}/dr)_i}{ \sum_{i=i_{\rm min}}^{i_{\rm max}}(dA_{Ks}/dr)_i}
\end{equation}

The extinction is converted to a column density using the empirical relation 
between {\changes the extinction in the $J$ band,} $A_J$, and N$_{\rm H}$ found by \cite{Vuong2003} in $\rho$ Oph dark cloud: 
\begin{equation}
  \label{Vuong}
  N_{H} = 5.57 \times 10^{21} A_J\, {\rm cm^{-2}} \, {\rm mag^{-1}}
\end{equation}
{\changes where $N_{H}$ is the column density of hydrogen} derived from observations of x-ray absorption.  
{\changes We have chosen this value over the more well known \cite{Bohlin1978} relation 
( $N_{H} = 1.9 \times 10^{21} A_V\, {\rm cm^{-2}} \, {\rm mag^{-1}}$) as the \cite{Vuong2003} result was obtained 
in a dense star forming region, more 
similar to the environment that we expect to find in IRDCs. It should be noted, however, that the impact on our mass determinations 
between the two is only around $\sim 20\%$}.
As our extinction results are in the $Ks$ band, we adapt the \cite{Vuong2003} relation using the  {\changes \cite{Cardelli1989} extinction curve ($A_V : A_J : A_{Ks} = 1 : 0.282 : 0.114$)}.

In order to transform this into a mass, the column density is multiplied by the area of the cloud and 
a multiplicative factor of 1.36 \citep{Allen1973} is applied to take into account the presence of Helium and other heavy elements.
The mass of a cloud with area $\Sigma_{\rm IRDC}$ pc$^2$ is then :
\begin{equation}
  \frac{M_{\rm IRDC}}{M_{\odot}} = 150 \times \left(\frac{A_{Ks}}{\rm mag}\right) \times 
  \left(\frac{\Sigma_{\rm IRDC}}{\rm pc^2}\right).
\end{equation}
This method assumes that the cloud covers the entire ellipse defined in the \cite{Simon2006} IRDC candidate catalogue. 
As this is not always the case, this supposition may result in the mass being overestimated. On the other hand, 
if parts of the cloud completely obscure the NIR light emitted by background stars then the cloud mass will be underestimated.
Nevertheless, the resulting mass is expected to be accurate to within a factor of a few.

\section{Results}
\label{sec:results}
The present method has been applied to over 1500 IRDC candidates (using the selection described above) 
and we have detected 1259 of them with an $A_V \ge 1$.
{\changes
The use of the genetic algorithm allows us to apply the technique to a lower number of stars than the method published in  \cite{Marshall2006}, allowing 
us to probe smaller and denser clouds. As a result, the current method results in a factor of four more cloud detections than the older method.
}
Due to the number of detections, the full table of the IRDC characteristics 
will only be available electronically, 
via the VizieR service\footnote{http://vizier.u-strasbg.fr/viz-bin/VizieR}, but an example of the results 
is shown in table 1.
The distribution of the detected IRDCs in the plane of the Galaxy is displayed in Fig.\ref{fig:irdc_pos}.       
The overlaid spiral arms are from \cite{Vallee2008}, with the addition of the local Orion spur. 
The bar angle plotted here is 20$\degr$, in agreement with recent values \cite{Gerhard2002}. This spiral structure is
 not constrained by the distribution of IRDCs presented here, but is overplotted for comparison.
The clouds are seen to preferentially lie along the spiral arms 
of the Galaxy, as well as in the molecular ring ($3<R<6$). Also, a
 concentration of IRDCs is visible at the end of the near side of the Galactic bar.
These regions are all associated with high rates of star formation,
  reinforcing the idea that IRDCs are the precursors to massive stars.

\begin{table*}
  \begin{center}
    \begin{tabular}{lccccccccc}
      \hline
      \hline
      Name & $l$ & $b$ & Distance & $A_{Ks}$ & Size & Mass &$N_{\rm obs}$ & $\chi_r^2$ & $\sigma$ \\
      & \degr & \degr & kpc & mag & pc & $M_{\odot}$ & & & \\
      \hline
      G025.65+00.55 & 25.6550 &  0.5590 & 3.92 & 0.580 & 8.94 & 5768 & 381 & 2.43 & 51.55\\
      G025.66-00.12 & 25.6690 & -0.1260 & 4.69 & 0.160 & 4.57 & 416 &  60 & 0.60 & 14.11\\
      G025.72-00.28 & 25.7240 & -0.2870 & 4.85 & 0.200 & 4.64 & 535 &  65 & 0.80 & 32.41\\
      G025.74-00.44 & 25.7450 & -0.4410 & 4.35 & 0.180 & 13.43 & 4044 & 726 & 3.11 & 9.90\\
      G025.79+00.81 & 25.7990 &  0.8150 & 4.10 & 1.040 & 5.96 & 4594 & 116 & 1.26 & 67.43\\
      G025.85+00.49 & 25.8530 &  0.4950 & 3.98 & 0.900 & 6.07 & 4134 & 102 & 1.01 & 439.74\\
      G025.90+00.33 & 25.9060 &  0.3350 & 4.29 & 0.470 & 5.17 & 1563 & 100 & 0.68 & 41.12\\
      G025.92+00.26 & 25.9260 &  0.2660 & 7.06 & 0.170 & 9.38 & 1864 &  92 & 0.67 & 3.79\\
      G025.94+00.63 & 25.9420 &  0.6390 & 2.63 & 0.280 & 3.09 & 332 & 134 & 0.94 & 11.34\\
      G026.18+00.14 & 26.1900 &  0.1500 & 5.07 & 0.460 & 5.20 & 1547 &  74 & 0.97 & 22.18\\
      \hline
    \end{tabular}
    \caption{Example of the results for the IRDC characteristics. For each cloud is listed, in order, it's name, 
      Galactic coordinates $l$ and $b$, distance, extinction, size, mass, the number of 
      2MASS stars used in extinction calculation, the reduced $\chi^2$ (Eq.4) and the 
      number of standard deviations the density of the cloud sits above the background (see Fig.3).}
  \end{center}
\end{table*}

\begin{figure}[t]
\begin{center}
    \includegraphics[width=0.8\linewidth]{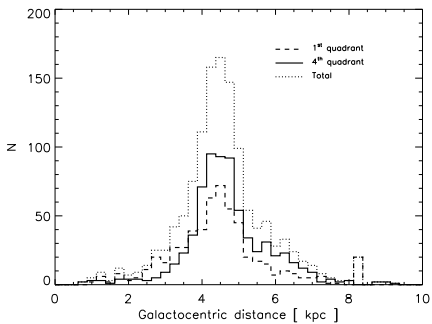}
    \caption{Distribution of Galactocentric distances of IRDCs: Dotted line (all clouds), dashed line(1$^{\rm st}$ quadrant) and solid line (4$^{\rm th}$ quadrant). 
The peak occurs in the molecular ring.}
    \label{fig:irdc_dist_histo}
\end{center}
\end{figure}

The distribution of the clouds in Galactocentric distance is shown in Fig.\ref{fig:irdc_dist_histo}. 
Few clouds are seen to lie within 
3 kpc of the centre. The bulk of the clouds are concentrated between 3 and 6 kpc, corresponding to the molecular ring. 
This is reflected in both the 1st and 4th quadrant IRDCs, however there is an asymmetry at $R=6$ kpc due to the presence of the 
Carina arm. Furthermore, the local Orion spur is present in the 1st quadrant only, creating a peak at $R\sim8.5$ kpc.
There are more IRDCs in the 4th quadrant, however this may just be due to the presence of more spiral arm tangents in this quadrant, making the detection of IRDCs easier.

The range of masses of the detected IRDCs 
run from {\changes just over ten solar masses up} to $8.7\times 10^4 M_{\odot}$, with an average mass of  $\sim 3.5\times 10^3 M_{\odot}$.
Defining their size as the radius of 
a circle with the same area as the IRDC ellipse, the average size of the IRDCs is $\sim 7$pc with the size ranging from 
a minimum of 0.9 to a maximum of 64 pc.
The average extinction is $A_{Ks} \sim 0.35$ mag corresponding to a total hydrogen column density of $\sim 5\times 10^{21}$cm$^{-2}$ while 
the maximum extinction detected is  $A_{Ks} = 1.32$ mag or $\sim 2\times 10^{22}$cm$^{-2}$. As pointed out by \cite{Simon2006a}, this range places IRDCs 
somewhere between the larger, more diffuse
 giant molecular clouds, and the smaller, denser Bok globules.

The mass spectrum of the IRDCs is estimated by binning the mass 
distribution into logarithmically spaced bins. 
The mass spectrum is then {\changes calculated from} the number of clouds in each bin divided by the size of the bin :
\begin{equation}
  f(M) = \frac{dN}{dM} \approx \frac{N_i}{\Delta M_i} 
\end{equation}
where $N_i$ is the number of clouds in bin $i$  and $\Delta M_i$  is the size of the $i^{\rm th}$ bin.
Mass spectra constructed using binned masses can introduce systematic errors as the 
slope is sensitive to the choice of the bin size. However, for samples 
with $N>500$ the resulting uncertainty in the spectral index is lower than 0.1 \citep{Rosolowsky2005}.

\begin{figure}[t]
\begin{center}
    \includegraphics[width=0.8\linewidth]{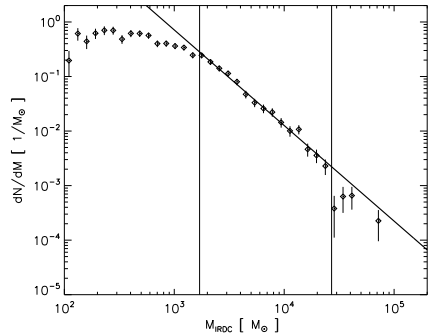}
  \caption{Mass spectrum of the IRDCs. After $\sim 1.7\times 10^3 M_{\odot}$, the spectrum is well fit by a power law (solid line)
    {\changes with a possible break at high mass. The two vertical lines delimit the zone where the cloud sample is deemed to be complete.}}
  \label{fig:irdc_mass}
\end{center}
\end{figure}

The binned mass spectrum of the IRDCs is displayed in Fig.\ref{fig:irdc_mass}.
The errors shown do not take into account errors in the mass estimation, and are simply counting errors : $\sigma=\sqrt N / \Delta M$. 
Small clouds are undersampled, due to the relatively large amount of stars necessary to apply the 3D extinction technique, and 
to the fact that large clouds are much easier to detect at large distances. For masses greater than $\sim 1.7 \times 10^3$ $M_{\odot}$, 
the mass spectrum is seen to follow a power law ($f(M)\propto M^{\gamma}$).
A straight line fit to the spectrum for $M>1.7 \times 10^3 M_{\odot}$ 
yields a spectral index of $\gamma = -1.75\pm0.06$ (solid line in Fig.\ref{fig:irdc_mass}), steeper 
than reported values for molecular clouds in the inner Galaxy \citep{Solomon1987, Rosolowsky2005} and flatter than 
the slope found by \cite{Simon2006a} for their IRDC selection.
{\changes The data seem to suggest that at the highest masses there is a break in the power law. 
However the average number of clouds per bin at these masses is just over 2, so it is not possible to reach any firm conclusion. 
These points (to the right of the second vertical line in Fig.\ref{fig:irdc_mass}) have not been used in the linear fit
 One possible explanation to the high mass cutoff seen may be  high mass clouds being  mistakenly identified  as several smaller mass clouds
in IRDC input catalogue, an effect mentioned by \cite{Rosolowsky2005}. 
In order to tackle this problem, the IRDCs could be characterised using higher 
resolution MIR observations such as the GLIMPSE survey, which would enable a more robust characterisation of the IRDCs. }



{\changes 
\cite{Hennebelle2008} 
{\changesa examine} the mass spectral index $\gamma$ of a cloud population 
 governed by density fluctuations induced by turbulence. They 
find that $\gamma = -2 + (n'-3)/3$ where $n'$ is the power spectrum index of $\log(\rho)$. They also mention that 
simulations show 
$n'$ to be close to the Kolmogorov turbulence index, or $n'\sim 11/3$. This results in a mass spectral index of $\gamma \sim -1.77$ 
- very 
close to what we find. This suggests that the IRDCs in our sample are not virialised but are the result of density fluctuations 
induced by turbulence. {\changesa This observation does not exclude the presence of higher density, autogravitating
 regions within the individual 
clouds in our IRDC selection. These higher density regions could in fact be completely missed by our method 
 as the column density of the dust would be too 
high to observe background stars in the near infrared and at the 2MASS completeness limit.}}

\section{Discussion}
\label{sec:discussion}
\subsection{Limits of NIR extinction}

The main difficulties in using the present method to obtain information on the IRDCs are the small sizes and large densities of the clouds.
The technique we employ requires the detection of many stars along the line of sight to obtain enough information on the 3D extinction. This 
rules out small clouds and clouds so dense that too few stars are detected behind them.
Thus, for our initial selection, 
we start by rejecting all clouds that lie along lines of sight having fewer than 50 stars detected in the three 2MASS bands.

The completeness of the 2MASS catalogue is limited by source confusion, and is a problem in the Galactic plane and towards the Galactic bulge. 
However, as we are concentrating on high extinction features, source confusion is less of a problem and the 2MASS completeness limit 
is less severe. In fact, the average completeness in three bands of our IRDC sample is 15.2, 13.9 and 13.1 for the $J$, $H$, and $Ks$ 
bands, respectively. As we use a bright cutoff at magnitude 10, the largest $J-K_s$ observable is 5.2, which results in a maximum 
extinction of $\sim 3.5$ in the $K_s$ band and a maximum column density of $5.1\times10^{22}$ cm$^{-2}$, from Eq. \ref{Vuong}. 
The minimum extinction detectable 
results from the colour uncertainty in the 2MASS dataset, and {\changes corresponds to} about  $7\times10^{20}$ cm$^{-2}$. This range of densities is compatible 
with the typical densities found for IRDCs through  $^{13}$CO observations \citep{Simon2006a}.

\subsection{Comparison with existing measures}

\cite{Simon2006a} measured the $^{13}$CO line towards over 300 candidate IRDCs in the first Galactic quadrant. They 
used the radial velocity channels of the CO observations that matched the extinction features in the MSX data \citep{Simon2006}. 
They then attributed a kinematic distance to the cloud by assuming circular motion and using the \cite{Clemens1985} rotation curve. 
They solved the inner Galaxy distance ambiguity by assuming that all clouds are at their near kinematic distance as 
the clouds are seen in extinction and therefore must lie in front of the Galactic background emission. The masses were calculated 
from the CO emission and by assuming an H$_2$ to CO conversion factor. 
\begin{figure*}[t!]
  \begin{center}
\includegraphics[width=0.3\linewidth]{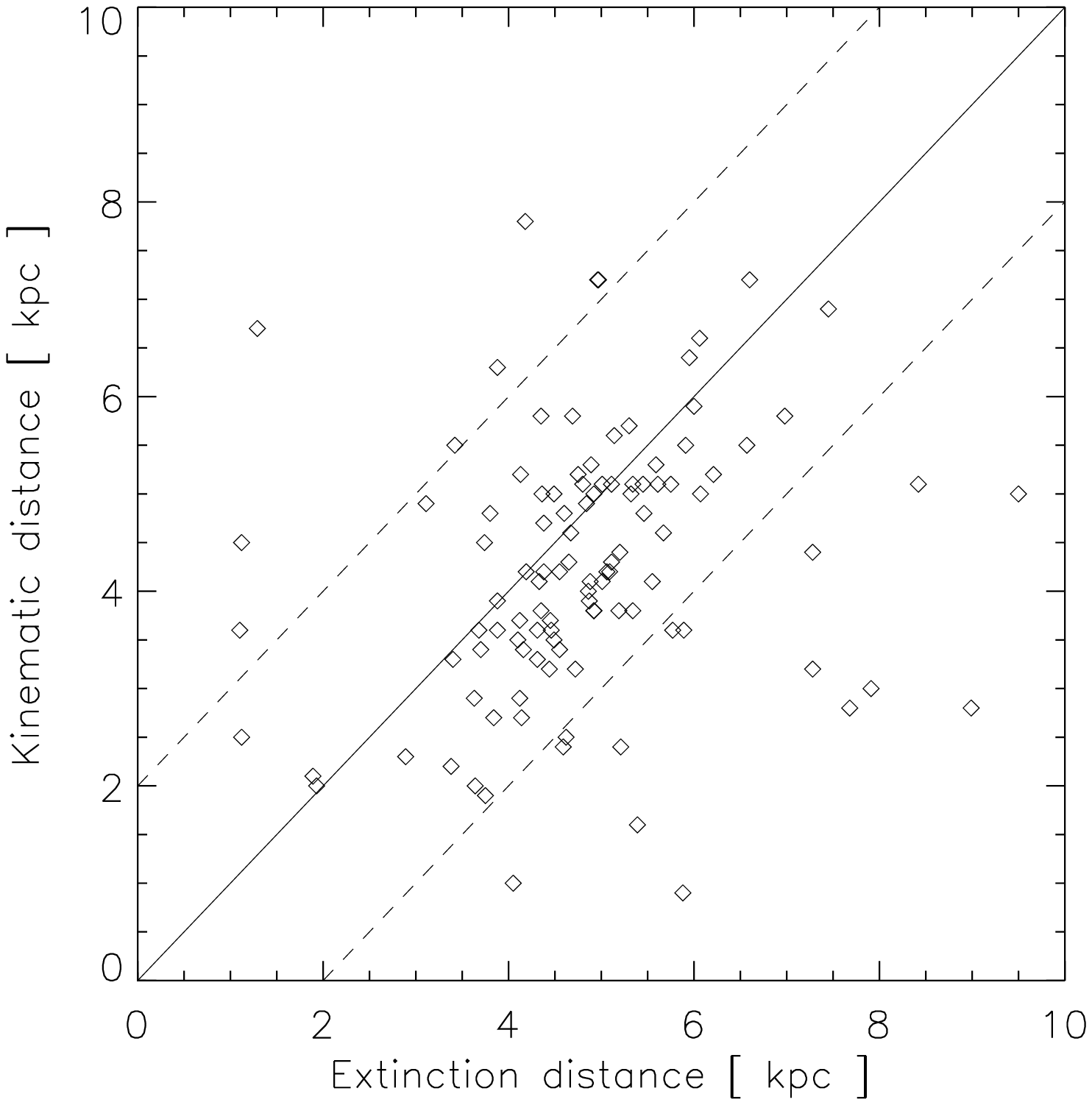}
\includegraphics[width=0.3\linewidth]{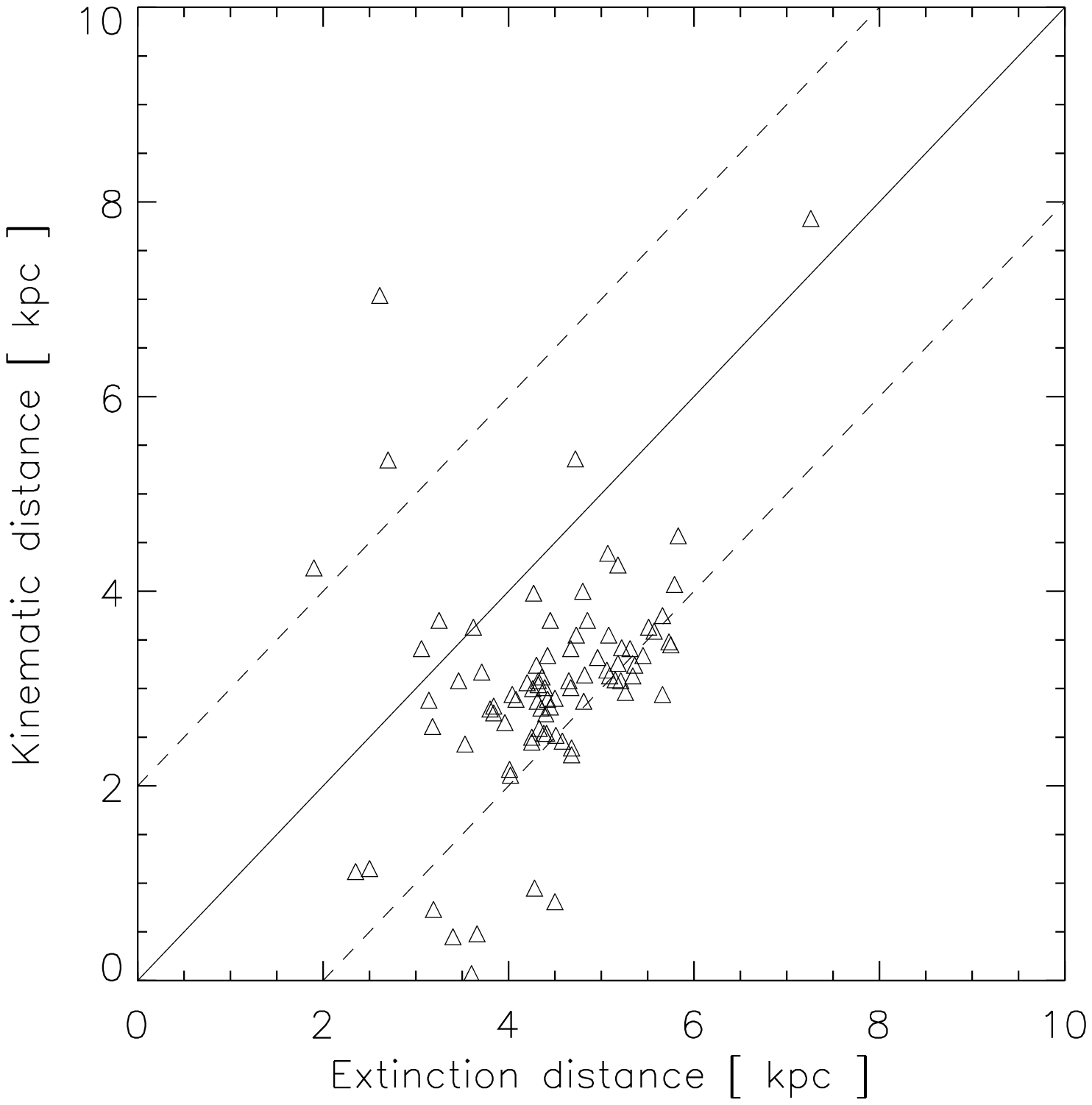}
    \includegraphics[width=0.3\linewidth]{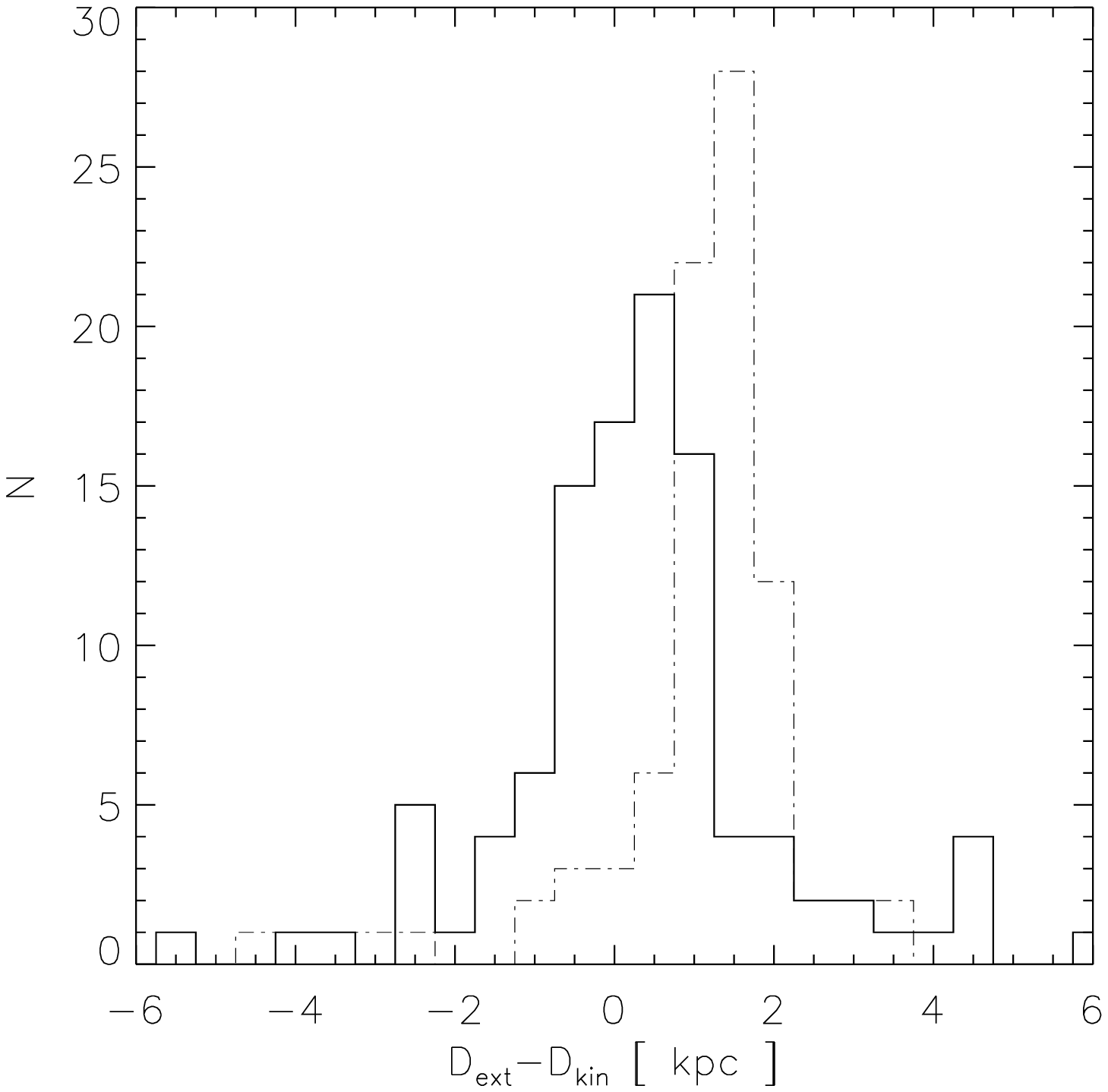}
\caption{{\bf Left} : Scatter plot of extinction distances with the kinematic distances listed in \cite{Simon2006a}. The black line shows equality 
  and the dotted lines are at $\pm 2$ kpc.
  {\bf Middle} : Comparison of extinction distances with the kinematic distances listed in \cite{Jackson2008}.
    {\bf Right} : Histogram of differences between extinction distances and kinematic distances as found by \cite{Simon2006a} (solid line) and \cite{Jackson2008} (broken line). 
    A significant $\sim 1.5$ kpc systematic difference is found with the distances reported by \cite{Jackson2008}.}
  \label{fig:simon_comp}
  \end{center}
\end{figure*}


{\changes We have detected 107 of their sample  using the present method} - the other clouds in their study 
either have too few 2MASS stars detected towards them ($>75\%$ of their sample) or do not present an extinction feature 
with $A_V > 1.$ For a number of their lines of sight, they include 
distances for two or more line of sight clouds; we take the highest density cloud to compare with our estimate.
The different results for the distances of 
this IRDC sample are compared on the left side of Fig.\ref{fig:simon_comp}. The black solid line shows equality and the dotted lines are placed at 
$\pm 2$ kpc from equality. Although there is appreciable scatter, there is general agreement for most of the points. 
Clouds for which extinction distances differ greatly from the kinematic distances may, in fact, be different clouds. As we have not mapped out the cloud, we 
are not able to morphologically match the extinction in the NIR to the MIR extinction features as \cite{Simon2006a} do. On the other hand, some of their 
distances estimates may be wrong due to their systematic choice of the near kinematic distance.
On the right hand side of the Fig.\ref{fig:simon_comp} is the histogram of the difference between the two distance determinations. Here 
we can see that over 80$\%$ of the distance determinations are within 2 kpc of each other. There is only a slight systematic 
offset of $\sim$0.5 kpc.

\cite{Jackson2008} obtained distances for IRDC candidates from the \cite{Simon2006} sample in the fourth quadrant using 
observations of the CS molecule. This molecule requires high density for excitation making it a good tracer of dense clouds. 
Like us, they do not map the clouds but search for the highest density feature along the line of sight.
In the centre of Fig.\ref{fig:simon_comp} we compare our distance measurements to theirs, {\changes for the 95 lines of sight where 
both methods detect a cloud}. 
For a number of their lines of sight, they include 
distances for two line of sight clouds; we take the higher density cloud to compare with our estimate. There is a systematic offset of around 1.5 kpc 
between the two methods (Fig.\ref{fig:simon_comp}, right), with just a few points differing by much more. As mentioned above, these outliers may be due to different line 
of sight clouds being detected with the two methods or may be due to the systematic choosing of the near kinematic distance by \cite{Jackson2008}.

\begin{figure}[t!]
  \begin{center}
    \includegraphics[width=6cm]{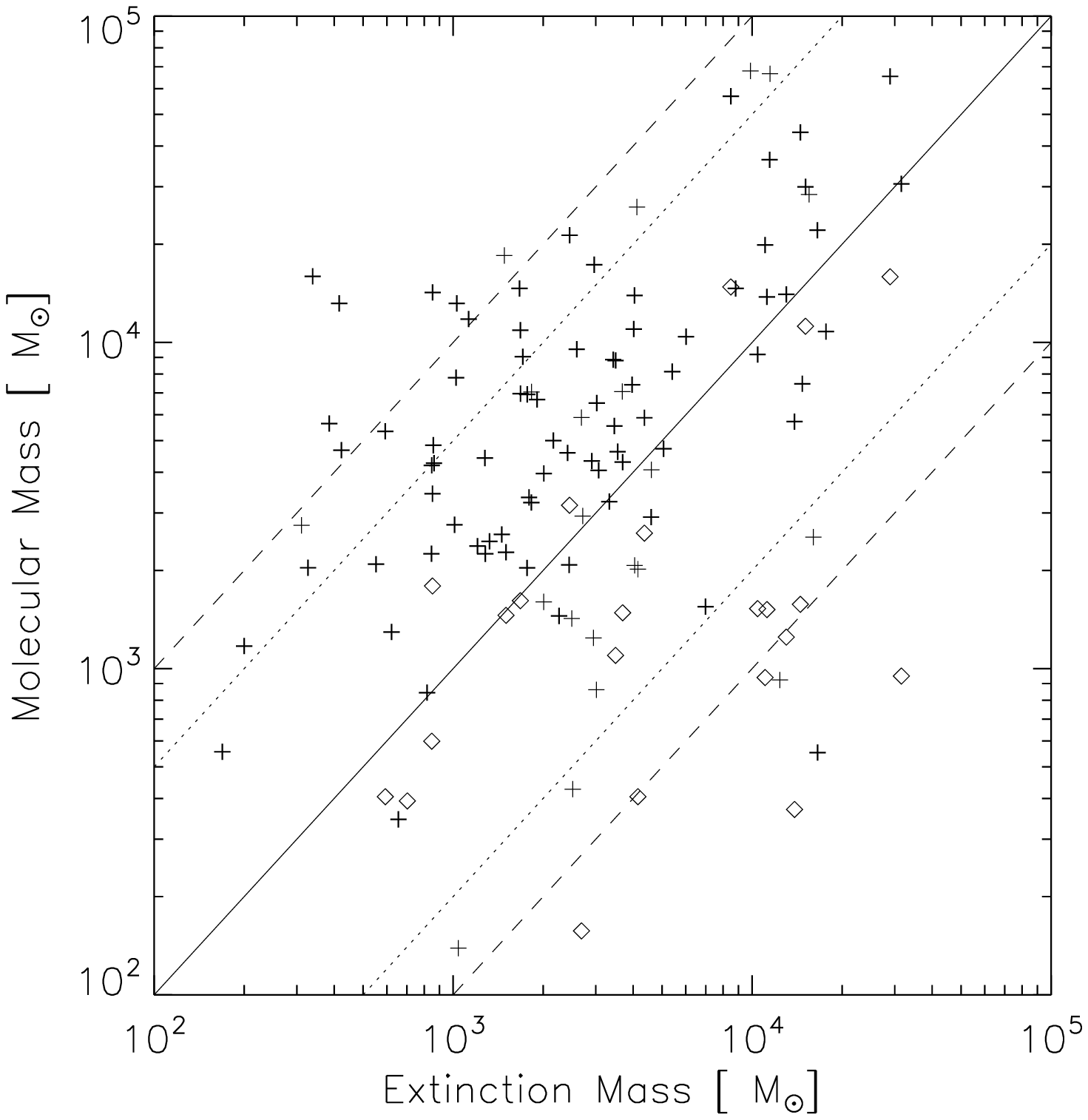}\\
    \caption{ Comparison of our mass estimates with those from \cite{Simon2006a}(plus signs) and \cite{Rathborne2006}(diamonds).}
  \label{fig:simon_mass}
  \end{center}
\end{figure}

As the mass depends on the square of the distance, the latter is key in determining accurate masses. 
In Fig. \ref{fig:simon_mass} we compare our 
mass estimates with those from \cite{Simon2006a} ($+$) and \cite{Rathborne2006} ($\diamond$). 
Note that Fig. \ref{fig:simon_mass} is a log-log plot. 
There is appreciably more scatter here {\changes compared to that observed for the distances. Most mass estimates agree within a factor of ten and}  
our mass estimates are generally larger than the \cite{Rathborne2006} estimates.
Considering the uncertainties in the three methods, an order of magnitude difference between methods is to be expected {\changes as the techniques used trace 
different parts of the IRDCs}. 

{\changes 
CO observations of IRDCs are a good tracer of the lower density envelope, but may be optically thick towards the denser cores. Also,  
 the low temperatures in the cores {\changesa may cause the CO molecules to form} 
ice mantles on the dust grains \citep{Alves1999}. Observations of the dust mm-continuum is optically thin, even at the very high column densities 
of the IRDC cloud cores. On the other hand, current mm observations are more sensitive to the small scale structure but miss out much of the extended 
emission of the cloud \citep{Bergin2007}.
}
Thus it is not surprising that our mass estimates are closer to those of \cite{Simon2006a}, as our method will also 
miss out the dense cores while recovering the lower density envelope, and that they are higher than \cite{Rathborne2006} as most of the mass of dark clouds 
is contained in the lower density envelope \citep{Alves1999,Cambresy2002,Lombardi2006}.


By comparing their results with those from \cite{Simon2006a}, \cite{Jackson2008} report an asymmetry 
in the distribution of 1$^{\rm st}$ quadrant IRDCs compared to 4$^{\rm th}$ ones. Our analysis does not show this asymmetry, however 
we find that the kinematic distances in the first quadrant have less of an offset to the extinction {\changes distances} than the fourth quadrant 
IRDCs. This may be evidence of a difference in the velocity field between the two quadrants, or may be due to a large scale stellar asymmetry 
not present in the Galactic model. Indeed, results from the Galactic Legacy Mid-Plane Survey Extraordinaire \citep[GLIMPSE][]{Benjamin2005} 
show stellar overdensities along a number of lines of sight crossing spiral arm tangencies. However these are present in both the first and fourth 
quadrants with similar intensities, and so they do not seem to be responsible for introducing 
such a large bias in our measurements. Furthermore, a number of precise parallax measurements of star forming regions using the 
NRAO Very Long Baseline Array (VLBA) show 
very good agreement with the kinematic distances towards massive star forming regions 
in the first quadrant \citep{Brunthaler2008,Zhang2008}, although, to our knowledge, no similar
measurements exist for the fourth quadrant.
Kinematic distances in the region studied have uncertainties of 0.5-1.5 kpc and occasionally up to 3 kpc 
\citep{Gomez2006} due to non-circular motions associated with the spiral arm shocks. The distance obtained 
from the 3D extinction has an uncertainty of $\sim 0.5-1.$ kpc. However, 
the selection of IRDCs seen in extinction may be polluted by other line of sight clouds. 

\subsection{Galactic spiral structure}

The distribution of IRDCs shown in Fig. \ref{fig:irdc_pos} is seen to follow the spiral structure of the 
Galaxy. However, departures are to be seen, especially between the Norma and Scutum arms at Galactocentric 
distances of between 4 and 5 kpc. This distance matches that of the Molecular ring \citep{Taylor1993,Dame2001}
which may be a ring like structure or may be the result of current angular resolution being unable to resolve 
multiple spiral arms in this region \citep{Vallee2008}. The {\changes spiral pattern traced out by the IRDCs 
in Fig. \ref{fig:irdc_pos}} could be explained by either an actual molecular ring or the bifurcation of one arm into two. 
Kinks and wiggles are present in every external spiral galaxy observed, so we are not to expect perfect 
logarithmic spirals from our own host Galaxy. The current results, however, are not enough to refute 
or confirm the molecular ring hypothesis, as there may be some (as yet) unknown bias affecting the results in the 
4th quadrant. 

{\changes 
The current catalogue of IRDC candidates \citep{Simon2006} 
{\changesa is not a complete survey of cold dark, ``IRDC-like'', clouds in the Galaxy. Indeed, it}
has been compiled with a severe bias - as the clouds are seen in extinction in the MIR, 
they require a bright background to be detected. As such they are almost exclusively on the near side of the Galaxy and are 
concentrated at very low Galactic latitudes. With the recent launch of Planck, a full sky survey of sub-millimetre emission 
from cold dust will soon be accessible. These observations will provide an unbiased survey of the cold cores which 
are known to be associated with IRDCs. The present 3D extinction method could be adapted to probe these {\changesa higher 
density} cores by utilising 
{\changesa stellar} MIR observations, such as the GLIMPSE survey (see Fig.\ref{fig:2mass_glimpse}), 
or the upcoming all sky MIR survey ``Wide-field Infrared Survey Explorer'' \citep[WISE,][]{Duval2004}. 

Compared to kinematic distances, the 3D extinction method has the advantage of providing temperature independent measures of the dust column density, as well 
as avoiding the near - far distance ambiguity which plagues kinematic distance estimates within the solar circle.
By using the MIR observations mentioned previously, along with the deep NIR observations of the Galactic Plane from the "UKIRT Infrared Digital Sky Survey (Galactic Plane Survey)" 
\citep[UKIDSS(GPS),][]{Lucas2008}, it will be possible to map out the spiral structure of our Galaxy to the far side of the solar circle and perhaps beyond.
}


\section{Conclusions}\label{sec:conclusion}

We have applied a new three dimensional extinction technique towards over 1500 IRDC cloud candidates, 
for which we recovered distances and masses for 1259 of them, including over 1000 previously unmeasured clouds.
This is a much larger sample than previous studies, enabling the study of the cloud population as a whole. 
The spatial distribution of the clouds is found to be concentrated in the molecular ring and 
along the spiral arms, reinforcing the idea that IRDCs are the birthplace for high mass stars. 
Their mass spectrum follows a power law with a slope of $-1.75\pm0.06$ which is steeper than GMCs in the inner Galaxy,
 similar to CO clumps and shallower than molecular cores or the stellar IMF.  
This slope suggests that the IRDC population is composed of non-gravitationally bound clouds, and are the result 
of density fluctuations induced by turbulence. {\changesa However,
 higher density autogravitating clumps inside the individual clouds of our IRDC selection would most likely
be missed by the present method.}

This new method is independent of any kinematical information, thus providing a new way to obtain information on 
the Galactic distribution of the ISM.  It is a good complement to existing measures which are solely based on 
molecular gas kinematics as both methods are completely independent and both are affected by different systematics. 
It will be able to provide valuable distance information for use in the 
analysis and interpretation of far-infrared and sub-millimetre observations by Herschel and Planck.
In the future it could be used with deeper stellar observations or observations at longer wavelengths 
in order to probe even higher density clouds and to even larger distances.

\acknowledgements
We would like to thank P.Hennebelle, G.Chabrier and M.Miville-Deschenes for helpful discussions. The anonymous referee provided 
a detailed report which helped improve the quality and clarity of the text.
D.J.Marshall was funded by the Natural Sciences and Engineering Research Council of Canada
through its SRO programme.
This publication makes use of data products from the Two Micron All Sky Survey, which is a joint project of the 
University of Massachusetts and the Infrared Processing and Analysis Center/California Institute of Technology, 
funded by the National Aeronautics and Space Administration and the National Science Foundation. 
The CDSClient package was used for the remote querying of the 2MASS dataset.
{\it Facilities:} \facility{2MASS}


\end{document}